\documentstyle[aps,prb,preprint]{revtex}

\begin{document}

\draft

\title
{ \bf \Large{
 Vortex Structure Around a Magnetic Dot in Planar Superconductors
}}

\author{
I. K.Marmorkos, A. Matulis \cite{adres} and  F. M. Peeters \\
}
\address{
Department of Physics, University of Antwerp (UIA), \\
B-2610 Antwerpen}

\date{ \today}

\maketitle

\begin{abstract}

The problem of the giant vortex state around a magnetic dot which is
embedded in a superconducting film is investigated.
The full non-linear, self-consistent Ginzburg-Landau equations are solved
numerically in order to calculate the free energy,
the order parameter of the host superconductor, the internal magnetic
field due to the  supercurrents, the corresponding current density,
the magnetization probed in the vicinity of the
dot, and the normal electron density
as a function of the various
parameters of the system.
We find that, as we increase the magnetic moment of the dot, higher
flux quanta vortex states  become energetically more favorable, as they can
better compete with the external magnetic field via the Meissner effect.
In addition to that, they progressively become closer to each other in
energy with direct experimental consequences, i.e. physical quantities
like magnetization may fluctuate when measured, for example, as a function
of a uniform external magnetic field.

\end{abstract}



\section{Introduction}
It is  well known that a crucial factor
determining the usefulness of a superconductor in practical applications  is
the maximum current at which it can operate (critical current). This current
is very small for type I superconductors.
In type II superconducting materials, however, large critical currents have
been measured which makes them more favorable for applications in e.g.
superconducting magnets.
In those materials the magnetic field lines partially penetrate  the
superconductor and transform it into the Abrikosov state by forming a
hexagonal lattice of vortices \cite{abrik,degen}. The
magnetic field lines penetrate the core of each vortex where the material
is in the
normal state, whereas, the rest of the system remains superconducting. The
application of a bias voltage results in  motion of these vortices giving
rise to dissipation which is now the limmiting factor for the
largeness of  the critical current of the superconductor \cite{tinkh,kim}.

To get around the above problem, and substantially enhance the critical
current, it is required to pin  the Abrikosov lattice as
strongly as possible.
 Defects in the crystal of size in the order of the
superconducting coherence length $\xi$ are very effective in vortex
pinning \cite{hueb}.
Experimentally, several types of defects have been utilized so far in studying
vortex
pinning: e.g.  point defects \cite{larkin} and amorphous
columnar defects \cite{civale,buzd}
created after bombarding the superconducting material with high energy
ions.

Since a full control over pinning is desirable, artificially fabricated
submicrometer holes in superconducting films have been recently
studied experimentally \cite{moshalk}. For certain "matching" fields, where
the period of
the Abrikosov vortex lattice and that of the lattice of holes were
multiple of each other, a strong pinning of vortices  was found
which resulted
in a strong enhancement of the critical current and in sharp peaks in the
magnetization curves \cite{moshalk}.
An alternative route exploited
 by van Roy {\it et al.} \cite{roy}
  was to
grow a
lattice of magnetic dots made of $\tau - MnAl$ on top of the superconducting
film.
At temperatures close to the transition temperature $T_{c}$ a strong
increase in the magnetization was measured, when the dots were magnetized,
indicating an enhanced pinning of the fluxoids by the modulated magnetic field
of the dots.

In the present work, we focus on  the later system and study in detail via the
Ginzburg-Landau formalism how
the superconducting film  is  perturbed in the neighborhood of the magnetic
dots. The paper is organized as follows: In the following section II we
describe the model on which our study is based.
Section III  discusses  the  technicalities of the numerical
integration of
the non-linear Ginzburg-Landau equations. In  section  IV, we
present and discuss our results.
Our results are summarized in section V.

\section{Model}

To better understand
the behavior of this novel system, experimentally studied
 by van Roy {\it et al.} \cite{roy},  we start from the simplest
possible theoretical model that, we believe, captures the qualitative
aspects  of the physics involved.
We consider a  single magnetic dot of radius $R$ embedded in a  planar
superconductor occupying the  infinite $x-y$ plane and
characterized by a Ginzburg-Landau parameter $\kappa= \lambda / \xi$, and
thermodynamic critical field $H_{c}$. $\lambda$ is
the penetration depth  of the superconducting material in
question  and $\xi$ is its coherence length.
The only source of   external field applied to the superconductor is
provided by the
magnetic dot whose
magnetic moment $\vec{m}$  is directed along the positive
$z-$axis and which gives rise to a  vector potential
 which, on the $x-y$ plane,   takes the form
\begin{equation}
\vec{A}_{0}(\vec{r})=(H_{c} \lambda ) \frac{\chi}{2 \kappa}A_{0}(r)
\vec{e}_{\theta} \,\, ,
\end{equation}
where we introduced the dimensionless vector potential
\begin{equation}
A_{0}(r)=\left( \frac{2m}{ \chi^{3} } \right) \frac{4}{k}
\frac{1}{ \sqrt{r} }
\left[ \left( 1-\frac{ k^{2} }{2} \right) K(k)-E(k) \right] \,\, .
\end{equation}
Here and further we shall use the unit vectors of the polar coordinate
system: $\vec{e}_r$, $\vec{e}_{\theta}$ and $\vec{e}_z$.
In the above expression, as well as in the rest of this work, we express
distance in units of the radius of the dot $R=\chi \xi$ ($\chi$ is the
dimensionless radius of the dot in units of the superconducting
coherence length $\xi$),
the magnetic field in units of $H_{c}$, and the magnitude of the magnetic
moment $m$ of the dot in units of $m_{0}=H_{c}(\pi \xi^{3})$. $K(k)$ and
$E(k)$ are complete elliptic integrals of the first and second kind
respectively with $k=4r/(1+r)^{2}$.
We should notice that on the plane of the superconductor the corresponding
magnetic field
$\vec{H}_{0}(r)=\vec{ \nabla}$x$\vec{A}_{0}(r)$
points
along the negative z-axis, for this particular choice of the magnetic
moment $\vec{m}=m \vec{e}_z$ of the dot.
In order to limit the number of parameters we assumed an infinite thin
magnetic dot when we calculated the vector potential (2).

We consider the magnetic dot, on top of the superconducting plane, to be
made of a hard magnet of uniform magnetization and that the structure of
its internal
diamagnetic
currents, which gives rise to its macroscopic magnetic moment $m$,
is not affected by the possible presence of nearby
circulating supercurrents. That is,
we realize a magnetic dot with rigid magnetic properties which serves
only as a source for the external non-uniform field $A_{0}(r)$ on the
superconducting plane. The magnetic lines of this field penetrate
the plane of the superconductor normally, have a radial symmetry, and
decrease in
strength as we move away from the dot giving rise to a magnetic dipole
field ($H_{0}(r) \sim m/r^{3}$) at large distances. Near the vicinity
of the dot, however, the radial dependance of this
non-uniform field on the $x-y$ plane is more complicated, and is given by a
combination of
elliptic integrals (see Eq. (2)), whereas, at the edge of the disk is
logarithmically divergent, since both the superconductor and the magnetic
dot are realized on the same $x-y$ plane. An alternative way to create
the same  magnetic field is by using a circular
loop with the same radius $R$  carrying a current $I=mc/(\pi R^{2})$.
The advantage of the later system is that this externally
imposed magnetic field can be tuned  in a controlled way by
changing the current on
the  loop. The fabrication of such current loops should  be feasible
nowadays with the advances in nanolithographic techniques.

The physical properties of the superconductor under consideration are well
described by the Ginzburg-Landau theory \cite{abrik,degen} which reduces to
the equations \begin{equation}
\frac{1}{2m} \left( -i \hbar \vec{ \nabla}
-\frac{2e \vec{A} }{c}  \right) ^{2} \psi
+ \alpha \psi +\beta \left| \psi \right| ^{2} \psi=0 \, ,
\end{equation}
and

\begin{equation}
\vec{j}= \frac{e \hbar}{im} \left(
\psi ^{*} \vec{\nabla} \psi - \psi \vec{\nabla} \psi^{*} \right) -
\frac{4e^{2}}{mc} \psi^{*} \psi \vec{A}  \,\, .
\end{equation}
The first equation gives the order parameter $\psi$  and the second one
the  supercurrent (diamagnetic response)
of the superconductor. In the absence of any external fields,  the order
parameter takes the constant value $\psi=\psi_{0}$ which is
determined by the density $n_{s}$ of Cooper pairs in the system. The
second equation, which is nothing more than  the usual quantum-mechanical
expression for the current in an external field,
should be  coupled to the Maxwell equation
$\vec{\nabla}$x$\vec{H}= (4 \pi /c)\vec{j}$. For a complete and
consistent description of the properties of the superconducting plane
under the external field $\vec{A}_{0}(\vec{r})$ the two non-linear
Ginzburg-Landau
equations, coupled together with the above Maxwell equation, should be
solved self-consistently with the appropriate boundary conditions. For
a superconductor-insulator interface, the theory of Ginzburg and Landau
requires the supercurrent across the interface to vanish, that is
\begin{equation}
\left( -i \hbar \vec{ \nabla} - \frac{2e}{c} \vec{A} (r) \right)
 _{n} \psi =0 \, .
\end{equation}


For the sake of convenience we write the total internal vector potential
$\vec{A}(r)$ in the superconductor as
\begin{equation}
\vec{A}(r)=(H_{c} \lambda) \left( \frac{ \chi}{2 \kappa} \right)
\left[ A_{0}(r) + \frac{1}{r} \phi (r) \right] \vec{e}_{\theta} \,\, ,
\end{equation}
where the dimensionless function $\phi (r)$, is obtained from the
self-consistent solution of the Ginzburg-Landau equations, and is directly
related to the vector potential created by the internal
currents in
the superconductor \cite{fink}. The corresponding total magnetic field
$\vec{H}(r)=\vec{ \nabla}$x$\vec{A}(r)$
is given by
\begin{equation}
\vec{H}(r)=H_{c} \left( H_{0}(r)+ \frac{1}{2r} \frac{d\phi}{dr} \right)
\vec {e}_{z} \, ,
\end{equation}
where the second term on the right hand side of the above equation is
the magnetic field
created by the supercurrents.

The radially symmetric magnetic field $H_{0}(r)$  created by
the magnetic
dot gives rise to superconducting vortices
with size  determined by the size of the dot. These vortices
  correspond to circulating
currents around the dot and form
 the so called
 ``giant vortex state'' \cite{fink,dela} which is closely to related the
superconducting
surface state. It  differs from the mixed state since it can carry
a total current, whereas the ideal Abrikosov state (without pinning
centers) cannot \cite{klose,abrikos}. Because of the circular symmetry of
our problem we take the order parameter
of the form
$\psi(r,\theta)=F(r)$exp$(iL \theta)$. The single-valueness of $\psi$ forces
the constant $L$ to be an integer. The correspondence of $L$ to the
orbital angular momentum quantum number, considering $\psi(r,\theta)$ to be
a wavefunction in the Schr$\ddot{o}$dinger-like Eq. (3), is evident,
and, in our system, it can be associated with the number of fluxoids
penetrating
the superconducting annulus region defined by the edge of the magnetic dot
and the circulating giant vortex.

A circulating current loop, with radius $r$ and current density
$\vec{j}(r)$, gives at $r+x$
 rise to a magnetic field $\vec{h} (r+x)$ such that
$|\partial h_{z} /\partial r| \, / \, |\partial h_{r} /\partial z| \sim 1/x$,
with $x \rightarrow 0$.
Therefore, for our particular problem of the giant vortex state, which
results from a circulating current density, we can safely reduce Maxwell's Eq.
$\vec{\nabla}$x$\vec{h}(r) =(4 \pi /c) \vec{j}(r)$
to
$-\partial h_{z} /\partial r = (4 \pi /c) j(r)$,
and write the two Ginzburg-Landau Eqs. in the final dimensionless form.

\begin{equation}
\frac{1}{r} \frac{d}{dr} \left( r \frac{dF(r)}{dr} \right) =
\left( \frac{ \chi^{2} }{ 2 \sqrt{2} \kappa } \right) ^{2}
\left[ A_{0}(r) + \frac{1}{r} \left( \phi (r)-B  \right) \right] ^{2} F(r)-
\chi^{2} F(r) \left( 1-F^{2}(r) \right) \, ,
\label{ginlan}
\end{equation}
and
\begin{equation}
\frac{d}{dr} \left( \frac{1}{r} \frac{d \phi (r) }{dr} \right) =
\left( \frac{ \chi }{ \kappa} \right) ^{2}
\left[ A_{0}(r) + \frac{1}{r} \left( \phi (r)- B \right) \right] F^{2}(r)
\, ,
\label{maxwell}
\end{equation}
with $B=(2 \sqrt{2} \kappa/ \chi^{2})L$.
In our model the magnetic dot serves {\em only} as
the source
of the external, non-uniform field acting on the superconductor which is
bounded by the magnetic disk.
Thus, we assume  hard wall boundary conditions
between the magnetic dot and the superconductor, and no magnetic field lines
due to supercurrents  are allowed to penetrate it.
If we notice that the magnetic field of a current loop is
much stronger close to the circumference of the loop, and drops fast,
as we move towards the center of the loop, we expext that this is not such
a bad approximation, which greatly facilitates our calculation.

Under these assumptions, the  boundary condition expressed by Eq. (5) is
translated into
$dF(r)/dr =0$ at $r=R$. From Eq. (6), however, it follows
that the vector potential $A_{s}(r)$ due to supercurrents is
$A_{s}(r) \sim \Phi (r)/r$ and, in accordance to our assumptions for the
properties of the hard wall at the edge of the dot,
  the usual definition of the magnetic flux
$\Phi_{s} =\oint \vec{A}_{s} d \vec{l} $ at $r=R$ gives $\phi(R)=0$.
 At large distances from the edge of the dot the normalized order
parameter $F(r)$ assumes the full value of the complete Meissner
superconducting state, that is $F(r)=1$, whereas, from the asymptotic
expansion of Eqs. (8) and (9) we find,  for $r \gg R$,
$\phi(r)-B \sim (2 \pi m/ \chi^{3})/r^{2}$. The corresponding field
$H_{s}(r)$ created by the supercurrents becomes
$H_{s}(r) \sim (2 \pi m /\chi^{3}) /(2r^{3})$, and exactly cancels
the field $H_{0}(r)$ of the magnetic dot. Matching the numerical
solutions of the above equations to these asymptotic expressions at
$r \gg R$  we naturally
recover the complete Meissner state of the plane superconductor at large
distances from the dot.

According to the above consideration we arrived to the following boundary
conditions
\begin{equation}
  dF(r)/dr=0|_{r=R}, \quad \phi(R)=0 \, ,
\label{start}
\end{equation}
and
\begin{equation}
  F(r)\rightarrow 1, \quad \hbox{and} \quad
\phi(r)-B \rightarrow (2\pi m/\chi^3)/r^2,
  \quad \hbox{when} \quad r\rightarrow \infty.
\label{fin}
\end{equation}

\section{Numerical integration}

We integrated numerically the system of two ordinary differential
equations (\ref{ginlan}), (\ref{maxwell}) with the boundary conditions
(\ref{start}) and  (\ref{fin}). This is a nonlinear two point boundary value
problem. We discovered that the integration cannot be done
straightforwardly,
and consequently, some kind of iteration technique should be used.

We constructed a superconvergent method which is based on the relaxation
technique (see, for example \cite{press}). The details of this method
are given in the Appendix, while the main result of it is the replacement
of the  nonlinear set of equations (\ref{ginlan}) and (\ref{maxwell}) by the
following linear set of equations  for the successive approximates
\begin{eqnarray}
\frac{d}{dr}F_{n+1} &=& \frac{1}{r}G_{n+1}, \nonumber \\
\frac{d}{dr}G_{n+1} &=&
\alpha_{00} F_{n+1} +\alpha_{01} \Phi_{n+1}
+ \beta_{0}  , \nonumber \\
\frac{d}{dr}\Phi_{n+1} & =& r\Psi_{n+1}, \nonumber \\
\frac{d}{dr}\Psi_{n+1} &=& \alpha_{10} F_{n+1} +\alpha_{11} \Phi_{n+1}
+ \beta_{1} \nonumber, \\
\label{iter}
\end{eqnarray}
with
\begin{eqnarray}
\alpha_{00} &=& r  \alpha \left(A_{0}(r) +\frac{\Phi_{n}(r)}{r} \right)^{2}
-r \chi ^{2} \left( 1-3 F^{2}_{n}(r) \right),
\nonumber \\
\alpha_{01} &=& 2  \alpha \left( A_{0}(r) +\frac{\Phi_{n}(r)}{r} \right)
 F_{n}(r) ,
\nonumber \\
\alpha_{10} &=& 2  \beta \left( A_{0}(r) +\frac{\Phi_{n}(r)}{r} \right)
 F_{n}(r) ,
\nonumber \\
\alpha_{11} &=& \frac{ \beta}{r} F^{2}_{n}(r),
\nonumber \\
\beta_{0} &=& -2  \alpha \left(A_{0}(r) +\frac{\Phi_{n}(r)}{r} \right)
F_{n}(r) \Phi_{n}(r)-2r \chi ^{2} F^{3}_{n}(r),
\nonumber \\
\beta_{1} &=& -  \beta \left( A_{0}(r) +2 \frac{\Phi_{n}(r)}{r} \right)
 F^{2}_{n}(r) ,
\nonumber \\
\label{iter}
\end{eqnarray}
where $\Phi_{n}=\phi_{n}-B$,  $\alpha =(\chi^{2}/(2 \sqrt{2} \kappa))$ and
$\beta =(\chi/ \kappa)^{2}$.
The boundary conditions (\ref{start}) and (\ref{fin}) are linear and they
remain the same for the above iterative scheme.

The obtained two point linear boundary problem is much more simple.
We integrated it straightforwardly using the standard supplementary
function technique (see, for example \cite{lance}). Namely, we integrated
the above nonhomogeneous linear equation set and the analogous homogeneous
equation set twice with the boundary conditions, and then
matched the result with the asymptotic function behaviour (\ref{fin}).
We took  particular care at the starting point ($r=R$)
where the vector potential $A_{0}(r)$ logarithmically diverges.
To properly handle this divergence, we used
the asymptotic expansion of Eqs. (\ref{iter}) to  analytically advance their
solution in the first step of the  integration.
We found out that the method was really superconvergent, and that
typically five to six iterations where sufficient.

\section{Results and discussion}

A crucial physical quantity that determines the behavior of the vortex
state around the magnetic dot, as a function of the various parameters of
the system, is the difference of the total magnetic free
energy  between the superconducting and the normal state
which is given by
\begin{equation}
\int \Omega _{SH} dV - \int \Omega _{NH} dV =
\frac{ H_{c}^{2} }{ 8 \pi}
\int \left(  \left( H(r) - H_{0}(r) \right)^{2}  - F^{4}(r) \right) dV \,.
\end{equation}

In our case, we are interested in the contribution of the giant vortex state
to the free energy, which we refer to the free energy of the
uniform superconducting plane (where the magnitude of that difference
acquires the constant value of $H_{c}^{2}/8 \pi$ per unit volume).
Therefore, we calculated the above integral over an annulus region of space
defined between $r=R$ and $r_{c}=20R$.
All the physics we are interested in (giant vortex state) takes place
within this region for all sets of parameters we investigated.
Our conclusions are not sensitive to the particular choice of $r_{c}$ as
long as $F(r_{c}) \simeq 1$ (i.e. we include all the region spanned by the
corresponding vortex state, whereas, the calculation of the above
integral from $r_{c}$ to infinity merely would add an unimportant constant).

 Relevant results
for the free energy difference  as a function of the
magnetic moment $m$ of the dot are shown  in Fig. 1
for  different quantum numbers $L$.
 Here, we consider a superconducting material
with a Ginzburg-Landau parameter $\kappa=1$ and a magnetic dot with radius
$R= \xi $.
The vortex state
with the quantum number $L$ which gives the lower free energy difference is
the one that is physically realized. In our system of a nonuniform
external magnetic field, we have
checked that such a state of a minimum total free energy gives  rise
to a free energy density which  is lowest everywhere in
space. This is shown explicitly in Fig. 2 for the $m=1$ case.
{}From Fig. 1 we see that for a particular value of $m=m_{c_{1}}$ the
complete
Meissner state ($L=0$) ceases to  be energetically the most favored one,
and a vortex state containing one flux quantum  appears. This value
of $m_{c_{1}}$  corresponds to the lower critical field $H_{c_{1}}$
we have in the
case of the {\em uniform} external field. A further increase of $m$
eventually introduces one more fluxoid in the system, and so on. From our
results we see that in the low $m$ regime the various vortex states
associated with
different flux quanta $L$  are energetically well
distinct from each other. At larger $m$ values, however, states
with different $L$ are very close together in energy, and, consequently,
the system can easily make transitions from one vortex state to another one
with a different quantum number $L$.
 This may result in
fluctuations when measuring various physical quantities of the system
(magnetization) as  a function of some external parameter, i.e. an
additional uniform, external magnetic field.
In fact, such peculiar fluctuations have been observed
experimentally \cite{roy}, and the mechanism we described could serve as a
guide of thought in trying to better understand them.
 This means that at large magnetic moments of the dot, where the
resulting field
is higher, the order parameter of the system is not a simple
function of one $L$-component only, but it should be written as a
superposition
\begin{equation}
\psi= \sum_{L} F_{L}(r) e^{iL \theta} \,\, .
\end{equation}
The larger the magnetic moment of the dot, the more $L$ components are
expected to have an important contribution in the above summation. In this
case,
the system  has a finite probability to be in any one of those states
characterized by a particular quantum number $L$. At low values of the
magnetic moment of the dot, however,
our simple model of keeping only one $L$ component is
more accurate, whereas, at larger $m$ values it still captures all the
qualitative aspects of the physics involved.

In Fig. 3 we present results for the order parameter $F(r)$ in the
superconducting plane as a function of distance from the edge of the magnetic
dot. Our results refer to a superconductor with $\kappa=1$, and to a dot
that has a magnetic moment $m=2$ and radius $R=\xi$.
The inset shows similar results but for $m=0$. For the above set of
parameters, we plot $F(r)$ for different values of the flux quantum
number $L$.
First of all, we see from the inset of Fig. 3 that for $m=0$, vortex
states with lower
$L$ values assume a higher value of the order parameter.
We have checked that the state with $L=0$ has the highest order
parameter, $F(r)=1$, everywhere in
space. From Fig. 1 we notice that this state gives a lower free energy
and energetically  is the most favored at $m=0$.
Close to the edge of the dot  the values of $F(r)$ for
different $L$ are well distinct from each other, and, within a
distance of about $5R$, $F(r)$ approaches unity, as  is the case for the
uniform Meissner state. The transition  of $F(r)$ to unity is rather broad,
and, the larger $L$ is, the larger the required distance for
$F(r)$ to reach unity.
At higher magnetic fields, (see Fig. 3) the state with $L=0$ has now a
drastically lower
order parameter which is a consequence of the fact that the $L=0$ vortex
state is no longer energetically favored (see Fig. 1). In the case of higher
fields, we notice that
the order parameters for different $L$ values are closer  to
each other, especially the ones for  larger $L$. The  reduction of the
superconducting state (lower $F(r)$) close to the edge of the magnetic
dot is more pronounced now, whereas, the transition of $F(r)$ to unity is
noticeably sharper  compared to the $m=0$ case.

Fig. 4 depicts  similar results for the order parameter $F(r)$, but
for a superconductor that has a larger Ginzburg-Landau parameter
$\kappa=3$ and a magnetic dot with $m=2$. We get  qualitative
similar behavior
with the one we had for the lower value of $\kappa=1$  at lower magnetic
fields, however.
For example, we have checked that the results for the order parameter
with $m=2$ and $\kappa=3$ show similar qualitative trends to the ones with
$m=1$ at $\kappa=1$. That is, increasing $\kappa$, the behavior
of the superconducting system scales at higher magnetic field values.
Thus,  our model recovers the well known fact that superconductors
with a
higher Ginzburg-Landau parameter $\kappa$ exhibit, at higher  magnetic
fields,
the same behavior which others (with lower $\kappa$ values) show at lower
fields, i.e. assume  higher critical magnetic fields \cite{abrik,degen}.

In Fig. 5 we plot results for the current density of the supercurrents as
a function of the distance from the magnetic dot with  $m=5$  ($m=1$ in the
inset) over different  values of the flux quantum number $L$ characterizing
the corresponding giant vortex state. First of all, notice that
the circulating
internal currents change
direction  at a particular distance  $r$, from
being negative to positive.
Thus, there are two currents circulating in opposite direction. This
phenomenon is similar to the behavior of bulk cylindrical
superconductors under a uniform external magnetic field \cite{fink}.
{}From our results of Fig. 5 we see that, for higher $m$,
states with
lower flux quanta give rise to  current densities that have a negative
sign of
circulation over larger distances, resulting finally in a total current
of negative sign. According to the  choice of our unit
vectors, a negative sign in the
internal current indicates that the
corresponding magnetic field created by the superconductor points
in the same
direction as the external field $H_{0}(r)$ of  the magnetic
dot. Such currents cannot compete and screen effectively the external
field, and are totally unable to give rise to the Meissner effect.
{}From Figs. 1 and 5 we notice that, for a particular value of $m$, flux
states with
extensive regions of negative
internal currents are energetically less
favored. For example, we see from the inset of Fig. 5 that for $m=1$ and
$L=2$ the
corresponding current density is positive over an extensive region of
space and gives
rise to a total current which is able to generate the Meissner effect.
At the same time (see Fig. 1) this state is energetically more
favored  to the  one with $L=1$ which lacks these features.
For $m=5$  states with  larger $L$ give rise to circulating currents
that better screen the external field and
more effectively minimize
the free energy of the superconducting system.
The $L=4$ state is the one with the lowest free energy for
$m=5$. If we conventionally
identify the positive peak of the current density as indicating an
effective radius $r_{v}$ of the system of the magnetic dot along with the
circulating
giant vortex, we find that $r_{v}$  decreases as the system
makes a transition to higher $L$ states.
According to our results in Fig. 1
these transitions take place more easily at higher $m$
values and finally results in
fluctuations of the effective radius $r_{v}$.
For $m=5$, $r_{v}$ could fluctuate between the states with $L=4$ and
$5$ resulting in fluctuations over distances up to $R/2$,
with direct experimental consequences. For example, the application of
an additional uniform external magnetic field would result in fluctuations
of the magnetization as a function of the magnitude of this field
\cite{roy}.

The circulating
internal currents around the magnetic dot, which form
the giant vortex state, give rise to a magnetic field
$H_{s}(r)$
whose  direction depends on that of the associated supercurrent.
The region in space over which the circulating current inverts direction
could also be a pronounced feature of the corresponding magnetic field
profile. To further investigate this,
we show in Fig. 6 the magnetic field $H_{s}(r)$ created  by these
circulating
currents as a function of the
distance from the edge of the dot for  $m=2$ for
different flux quantum numbers $L$. The inset shows similar results but
for $m=0$. First of all, we notice from the inset  for $m=0$ that
$H_{s}(r)$
decreases monotonously from the edge of the dot, being lower for lower
$L$ values, and approaching  zero at distances where the corresponding
order parameters approach unity. The picture for the  higher
$m=2$ field value, however, looks different in Fig. 6. The magnetic field
 $H_{s}(r)$
for low values of $L$ is negative for distances close to the
edge of
the magnetic dot, becomes positive at intermediate, and reduces to
zero at larger  distances from the dot.
A negative sign of the field $H_{s}(r)$ indicates that its direction is
the same as  the one of the  applied external field.
However, only when $H_{s}(r)$  is positive in sign can it compete with the
external field and give a reduced total internal magnetic field (Meissner
effect). From Fig. 6, however,  and for $m=2$ we see that
vortex states with a higher flux quantum number $L$ give rise to
a field   which is positive over a much larger distance,
and, along with their more effective competition
 with the  external magnetic field,
 they give rise to a lower free energy of the system.
We have, finally, checked that  the peak in $H_{s}(r)$ is exactly associated
with the point where the corresponding supercurrent inverts direction.
In Fig. 6 we plot also the external magnetic field $H_{0}(r)$
created by the magnetic dot in order to demonstrate  how this field is
exactly canceled at large distances  by $H_{s}(r)$
in order to recover the complete
Meissner state of the uniform planar superconductor.
{}From the energy diagram of Fig. 1 we see that the vortex state with $L=2$
is the one with the lowest free energy at $m=2$, and, according to our
results in Fig. 6, this state better cancels the external field $H_{0}(r)$
over a larger distance.

The internal
circulating  currents in the superconducting region around the
magnetic dot give rise to a magnetization
$4 \pi M(r)=H_{c}(H(r)-H_{0}(r))$.
The average value of it, in a finite region of space around the dot, can be
directly
probed experimentally using a SQUID device. A common outcome of such
experimental measurements is a hysteretic behavior of the magnetization, as
a function of the externally applied magnetic field \cite{saint},
indicating that the
superconducting system prefers to conserve the total number of flux quanta
trapped in it over a finite range of the external magnetic field.
Sometimes this may happen even at the cost of overriding the condition of
minimum free energy over a finite range of the external field
\cite{fink}. In Fig. 7 we show results for the  total
magnetization of a superconducting region spanned in a distance up to
$10R$, as a function of the magnetic moment $m$ of the dot over different
flux quanta numbers $L$. We notice the linear behavior of the
magnetization as a function of $m$. Transitions between lines of different
 $L$ take place at certain values of $m$. If the condition for minimum energy
dictates the behavior of the magnetization, such transitions take place
at the values of $m$ indicated by  the vertical dotted lines in Fig. 7.
Under
the condition of minimum energy requirement, the magnetization changes
reversibly. The same holds over a finite interval of $m$ such that the system
moves along a line of fixed $L$. If the system, however, prefers to
conserve the number of flux quanta at larger ranges of $m$, even at the
cost of overriding the condition for minimum energy imposed by Fig. 1,
then, a hysteretic behavior of the magnetization appears. In that case,
an increase of
$m$ over an appropriate, finite  interval, followed by a decrease back
to zero
leaves a certain number of flux quanta locked around the magnetic dot.

In Fig. 8 we plot
$n = \int_{R}^{\infty} (1-F^{2})rdr$,
 which is proportional to the total number of normal electrons in the
$x-y$ plane, as a function of  the magnetic moment $m$ over different
quantum numbers $L$. We see that, as we increase the externally applied
non-uniform  magnetic field (i.e. increase $m$), the superconducting
region around the dot progressively decreases.
This
shows up in Fig. 8 as an increase of the number of normal electrons
$n$ as we increase $m$.
After a careful comparison with the energy diagram of Fig. 1, we
note that this increase in $n$ always occurs for $L$ states that are
physically realizable (i.e. have a lower free energy) at the particular
value of $m$ we are considering.
We also notice that, at small values of $m$, $L$ states that
have the  minimum normal electron density are well separated from each
other.
At larger $m$, however, states with higher $L$, which according to the
results in Fig. 1 are energetically more favored, have normal electron
densities closer to each other.

\section{SUMMARY AND CONCLUSIONS}

As a summary, we studied the giant vortex state around a magnetic dot
embedded in a superconducting film. We found that at low values of the
magnetic moment $m$ of the dot vortex states with a low number $L$ of
flux quanta associated with them are energetically more favored and well
separated from each other in energy. At large $m$ values, however, vortex
states with a larger $L$ are now more favored, whereas,
they are  quite close together in energy. We pointed out that this may
result in fluctuating physical quantities when, for example,  we drive the
system with some uniform external magnetic field. In fact, such
peculiar fluctuations may be related to the ones which have been observed in
magnetization measurements
for a lattice of magnetic dots on  top of a superconducting film \cite{roy}.

\acknowledgements

Stimulating discussions  with V. V. Moshchalkov and W. Van Roy are
acknowledged.
This work was supported by the Belgian National Science Foundation,
 by NATO through the linkage grant: 950274, and by the PHANTOMS network
(ESPRIT Basic Research Action 7360).

\appendix

\section{ Superconvergent relaxation technique}

The nonlinear two point boundary problem is
a complicated numerical problem and can not be handled
straightforwardly. Here we present some general consideration which allows
to reduce it to a linear iterative scheme.
We shall illustrate our consideration by applying it to a first order
nonlinear differential equation set which we formally present as
\begin{equation}
\frac{d}{dx}f=H(f).
\label{basic}
\end{equation}
Here, the  symbol $f$ stands for some vector function with components
$f=\{f_1(x), \cdots , f_n(x)\}$, and $H(f)$ is some nonlinear function
of that vector. We shall also assume that  proper boundary conditions
accompany Eq. (\ref{basic}). The two point boundary conditions  are given
at the different points $x_i$.

The above equation set (\ref{basic}) is rather general, as
every system of ordinary differential equations can be
reduced to an equivalent set of first order differential equations.
Moreover, other equations (e.g., integral equation) can often be
presented in the form analogous to Eq. (\ref{basic}).

The main idea of the relaxation technique is the following. Instead of
solving equation (\ref{basic}) defined on the $x$-axis we shall consider
another equation
\begin{equation}
\frac{\partial}{\partial t}
     \left\{  \frac{\partial}{\partial x}f-H(f) \right\}= -
\left\{ \frac{\partial}{\partial x}f-H(f) \right\} \, ,
\label{relax}
\end{equation}
which is defined in the $xt$-plane.
When $t \to \infty$ its solution converges to the solution of
our basic Eq. (\ref{basic}).
For the sake of simplicity we shall assume that the boundary conditions
of our basic problem are linear and append them to Eq. (\ref{relax}).
We like to point out, however, that in the case of nonlinear
boundary conditions they can be handled in the same way, just replacing
them by the relaxation equations analogous to equation (\ref{relax}).

Now we shall replace the time derivative in Eq. (\ref{relax}) by the
approximate
expression
\begin{equation}
\frac{\partial}{\partial t}f \approx \frac{f(t+h)-f(t)}{h} \approx
\frac{f(t+1)-f(t)}{1}=f_{n+1}-f_n \,,
\label{der1}
\end{equation}
and
\begin{equation}
\frac{\partial}{\partial t}H(f)=\frac{\partial H}{\partial f}
\frac{\partial}{\partial t}f \approx \frac{\partial H}{\partial f}
(f_{n+1}-f_n) \,,
\label{der2}
\end{equation}
were $(\partial H/\partial f)f$ stands for
\begin{equation}
\frac{\partial H}{\partial f}f=\sum_{i=1}^n \frac{\partial H}{\partial f_i}
f_i(x).
\end{equation}

Finally, substituting expressions (\ref{der1}) and (\ref{der2}) into
Eq. (\ref{relax}) and restricting ourselves to
linear terms in $(f_{n+1}-f_n)$ only, we obtain the following set
of equations
\begin{equation}
\frac{d}{dx}f_{n+1}-H_{f}(f_n)f_{n+1}=
H(f_n)-H_{f }(f_n)f_{n} \,,
\end{equation}
which  is the required linear iteration scheme for our basic problem
(\ref{basic}).

Now denoting the derivatives of $F(r)$ and $\Phi(r)$ by $G(r)/r$ and
$\Psi(r)r$, respectively, and applying the above considerations to
Eqs. (\ref{ginlan}) and (\ref{maxwell}), we immediately arrive to the
iteration scheme given by Eqs. (\ref{iter}).

\newpage

\begin{figure}
\caption{
Free energy difference between the normal and superconducting state
($\kappa=1$) of
our system as a function of the magnetic moment $m$ of the dot for
various flux quanta numbers $L$.
$G_{0}=10^{3}(H^{2}_{c}/8 \pi)(\pi \xi^{2})$,
 $m_{0}=H_{c} (\pi \xi^{3})$, and the magnetic dot has a radius $R= \xi$.
 }
\label{fener}
\end{figure}

\begin{figure}
\caption{
Free energy density difference between the normal and superconducting
state ($\kappa=1$) of
our system as a function of distance from the edge
of the dot, which has a magnetic moment $m=1$, and radius $R=\xi$,
for various flux quanta numbers $L$.
 }
\label{dener}
\end{figure}

\begin{figure}
\caption{
Order parameter $F(r)$ of our superconducting system with $\kappa=1$
 as a function of the distance from the edge
of the dot, which has a magnetic moment $m=2$, (inset $m=0$), and radius
$R=\xi$, for various flux quanta numbers $L$.
 }
\label{orderf}
\end{figure}

\begin{figure}
\caption{
Order parameter $F(r)$ for the same system as in Fig.\,\,3, but now
for $\kappa=3$. \hspace{10cm}
 }
\label{orderk}
\end{figure}

\begin{figure}
\caption{
Current density  profile in the  superconductor
as a function of the distance from the edge
of the dot which has a magnetic moment $m=5$ (inset $m=1$) and radius
$R=\xi$ for various flux quanta numbers $L$. The corresponding
superconductor has a G-L parameter $\kappa=1$.
 }
\label{current}
\end{figure}

\begin{figure}
\caption{
Magnetic field profile generated by the supercurrents
as a function of distance from the edge
of the dot, which has a magnetic moment $m=2$ (inset $m=0$), and radius
$R=\xi$, for various flux quanta numbers $L$. The corresponding
superconductor has a G-L parameter $\kappa=1$. The external magnetic
field $H_{0}(r)$ due  to the magnetic dot is also shown.
 }
\label{field}
\end{figure}

\begin{figure}
\caption{
Total magnetization in a region defined between the edge of the dot and
$r_{0}=10R$
as a function of the dimensionless magnetic moment $m$ of the dot, which
has a radius $R=\xi$, for  various flux quanta numbers $L$.
$4 \pi M_{0}=10^{2}H_{c} (\pi \xi^{2})$. The vertical lines indicate the
transition points between the different $L-$ states as dictated by the
minimum of the free energy.
 }
\label{magnetz}
\end{figure}

\begin{figure}
\caption{
Number of normal electrons  $n$
as a function of the dimensionless magnetic moment $m$ of the dot, which
has a radius $R=\xi$, for various flux quanta numbers $L$.
 }
\label{density}
\end{figure}

\end{document}